\documentclass[12pt]{article}
\usepackage{graphicx}

\begin{document}
 
\author{
{\bf M. I. Mart\'{\i}nez, G. Herrera} 
\thanks{e-mails: mim@fis.cinvestav.mx, 
                 gherrera@fis.cinvestav.mx}\\
{\small Centro de Investigaci\'{o}n y de Estudios Avanzados} \\ 
{\small Apdo. Postal 14 740, M\'{e}xico 07000, DF, M\'exico}
}

\title{
Dalitz plot slope parameters for $K \rightarrow \pi\pi\pi$ decays
and two particle interference
}

\date{}
\maketitle

\begin{abstract}
We study the possible distortion of phase-space 
in the decays\\ $K \rightarrow  \pi \pi \pi$,
which may result from final state interference among the decay products. 
Such distortion may influence the values of slope parameters extracted
from  the Dalitz plot distribution of these decays. 
We comment on the consequences on the magnitude of violation of the  $\mid \Delta
I \mid = 1/2$  rule in these decays.
\end{abstract}

\newpage

%%%%%%%%%%

\section{Introduction}

%%%%%%%%%%

In most cases, strangeness changing hadronic decays obey the 
$\mid \Delta I \mid = 1/2$ rule.  Experimental information on the cases when 
the rule  is violated may help in understanding its origin. K-decays provide 
an important laboratory where one could learn more about this important 
issue \cite{weinberg}. In particular, by measuring
the slope terms of the $K \rightarrow 3\pi$ Dalitz plot, as suggested by Weinberg 
\cite{weinberg}, one can probe the $\mid \Delta I \mid = 1/2$ rule with good 
precision.\\

The slope parameters of the Dalitz plot distribution in non leptonic
$K_{3\pi}$ decays were measured some time ago with 
limited precision \cite{pdg}. More accurate experimental results
have been published recently \cite{new,batusov}, attracting renewed theoretical 
interest \cite{theory}.\\

The Dalitz plot distribution for $K \rightarrow 3\pi$
can be parametrized by a series expansion of the form \cite{pdg}
\begin{equation}
 \mid M \mid ^2 \propto 1 + g X + h X^2 + k Y^2 + ... 
\end{equation}
with 
$$
X=\frac{(s_3-s_0)}{m^2_{\pi}} \hspace{3cm}
Y=\frac{(s_2-s_1)}{m^2_{\pi}}
$$

\noindent
and 

$$
s_0=\frac{(m^2_K + m^2_{\pi_1} +  m^2_{\pi_2} +  m^2_{\pi_3})}{3} \hspace{2cm}
s_i=(p_K-p_i)^2
$$
\noindent
where $p_i$ are the four momenta of the pions ($i=1,2,3$) and the label 3 denotes 
the odd pion  in a decay.\\
The coefficients $g,h$ and $k$ are not available from theory, but can be extracted
from the experimental Dalitz distribution. In the literature there are other 
definitions of slope parameters.
We choose a definition compatible with the Particle Data Group \cite{pdg}.

%%%%%%%%%
 
\section{Bose-Einstein interference} 

%%%%%%%%%

Bose-Einstein Correlation (BEC) was observed for the first time by Goldhaber,
Goldhaber, Lee and Pais \cite{gglp} in $p\bar{p}$ interactions. They  found that
the angular  distribution of like-sign pion pairs was different from that of unlike-sign
pairs.
In general terms, the effect can be understood as the tendency of identical bosons
to occupy the same phase space. As a consequence, identical bosons are correlated
in their momenta, opening angles distributions, etc.

The impact of BEC on the Dalitz plot distribution in hadronic decays
of charm mesons has been studied before \cite{cuautle}. 
It has also been demonstrated to play an
important role in other phenomena \cite{residual} in high energy physics,
and on the extraction of standard-model parameters \cite{bialas, sjostrand}.
More recently, the effect of Bose-Einstein interference
on measurement of collective flow has been considered \cite{flow}.\\

In order for Bose-Einstein correlation to be present among identical
particles in the final state of any reaction, two aspects must be
fulfilled: i) the source of pions must be finite, ii) the emission must take
place chaotically to a certain degree.\\
In the cases under study {\it i.e.}, the decays:\\

\begin{equation}
K^- \rightarrow \pi^+ \pi^- \pi^-
\end{equation}

\begin{equation}
K^+ \rightarrow \pi^- \pi^+ \pi^+
\end{equation}

\begin{equation}
K^{\pm} \rightarrow \pi^{\pm} \pi^0 \pi^0
\end{equation}

\noindent
and 

\begin{equation}
K^0 \rightarrow \pi^0 \pi^0 \pi^0
\end{equation}

\noindent
condition i) is satisfied by the need of form factors to describe the
decay, and the well known finite size of the K mesons \cite{radius}.\\
In hadronic decays, such as those under study, particles are produced 
partially through  hadronization in a not a entirely coherent process.
The incoherence although small, we believe it to be sufficient to satisfy
condition ii), requiered for the presence of correlation.
Figure 1 illustrates the decay. The bubble represents the space-time region 
where particles are produced.

Correlations  among the decay products of a particle and the pions produced 
in the main  reaction, as described in Ref. \cite{residual}, will not be
considered here. We will rather regard the decay itself as a particle 
production process in which interference may arise.\\

The BEC are commonly described in terms of a two-particle correlation function:  
\begin{equation}
 R(p_1,p_2) = \frac{P(p_1,p_2)}{P(p_1)P(p_2)}
\end{equation}
\noindent
where $P(p_1,p_2)$ is the joint probability amplitude for the emission of
two bosons with momenta $p_1$ and $p_2$, and $P(p_1)$ and $P(p_2)$ 
are the single production probabilities.\\

The Bose-Einstein correlation among identical mesons has been used to 
probe the space-time structure of the intermediate state right before
hadrons appear \cite{gglp,us,boal} in high energy elementary particle and nuclear
collisions.\\

One parametrizes the effect assuming a set of point-like sources emmiting
bosons. These point like sources are distributed according to a 
density $\rho (r)$. The correlation function can then be written as, 
\begin{equation}
 R(\vec{p_1},\vec{p_2}) = \int \rho (\vec{r_1}) \rho (\vec{r_2}) 
\mid \psi_{BE}(\vec{p_1},\vec{p_2}) \mid ^2 d^3 r_1 d^3 r_2,
\end{equation}
\noindent
where $\vec{p_1},\vec{p_2}$ are the momenta of the two bosons, 
$\psi_{BE}$ represents the Bose-Einstein symmetrized wave function
of the boson system, with $\int _V \rho (\vec{r}) d^3 r = 1 $.\\

Taking plane waves to describe the bosons, one obtains
the correlation function for an incoherent source:
\begin{equation}
 R(\vec{p_1},\vec{p_2}) = 1 + \mid {\cal F} ( \rho(\vec{r}) ) \mid ^2,
\end{equation}
\noindent
where $\cal F (\rho)$ represents the Fourier transform of the density
function $\rho(\vec{r})$.\\

Phenomenological parametrizations of the effect have been proposed. 
to describe the quantum interference during the hadronization  in high 
energy reactions. For a recent review see Ref. \cite{boal}.  
Here, we use the GGLP \cite{gglp} parametrization to gain an idea of the impact
of particle interference on the slope parameter. In a more detailed study \cite{prep},
we will address other possible parametrizations.\\

The GGLP parametrization is one of the most commonly used. It is given by:
\begin{equation}
 R(\vec{p_1},\vec{p_2}) = 1 + \lambda e^{-\beta Q^2},
\end{equation}
where $Q^2$ is the Lorentz invariant, $Q^2 = -(p_1-p_2)^2$, which can be
written also as, $Q^2 = m^2_{12} - 4 m_{\pi}^2 $; here $m_{12}$ is the invariant
mass of the two pions and  $m_{\pi}$ the mass of the pion under consideration.
The parameter $\lambda$ lies between 0 and 1, and reflects the degree
of coherence in the production. The radius of the
source is defined by $r = \hbar c \sqrt{\beta} [fm]$.
The presence of Bose-Einstein correlations will modify not only the
invariant mass spectrum of like-charged but also that of unlike-charged pions. 
This reflection of BEC, known as residual correlation, 
has been studied so as to make sure that the reference 
sample of unlike-charged pions used to subtract the effect from the 
like-charged pions is free from any other correlations.\\
In particle production processes with much phase space high multiplicity, residual
correlations tend to be minimal. Nevertheless, some studies \cite{residual} claim
that this is not always the case, and that using unlike-charged pions as a reference
may be questionable.\\
In a particle production process yielding where only three particles
({\it e.g.}, $K \rightarrow \pi \pi \pi$)  residual correlations are 
expected to be significant.

%%%%%%%%%%%%%%%%

\section{Simulation of Bose-Einstein correlations}

%%%%%%%%%%%%%%%%

In this letter we want to estimate the effects of BEC on the phase space 
of a three body decay. In particular, on the phase space of the decay
$K \rightarrow \pi \pi \pi$. 
We will take the approach used in Ref. \cite{residual}, where
the BEC are simulated by simply weighting each event. 
The MC  generator consists of a three-body decay, with appropiate 
masses for the decay products. Each event is weighted according to:
\begin{equation}
W=\prod_{i,j} (1+ \lambda e^{-\beta Q_{ij}^2}),
\end{equation}
where the product is taken over all pairs $(i,j)$ of like-charged 
pions.\\
For cases when only two like-charged pions are present in the final state, as in 
the reactions (2-4), the  product reduces to
\begin{equation}
W= 1+ \lambda e^{-\beta Q^2}.
\end{equation}
However, for the decay (5) it becomes
\begin{equation}
W= (1+ \lambda e^{-\beta Q_{12}^2})(1+ \lambda e^{-\beta Q_{23}^2})(1+ \lambda e^{-\beta
Q_{13}^2}).
\end{equation}

\noindent
where the $Q_{ij}$ of all possible pairs are taken into account.
We performed simulations for different values of $\beta$ and $\lambda$.
The coherence parameter $ \lambda $ controls the strength of the effect, but does not
modify the shape of the distortion. 
The decay is not necessarily a completely chaotic process,
and $\lambda$ does not have to be 1. In fact, if
hadronization did not take place during
the decay, one would expect a completely coherent process in which $\lambda=0$, and
Bose-Einstein interference  would not be present. Hadronization in the
decay introduces some degree of incoherence giving rise to $\lambda$ values
between 0 and 1.
The exact value would be obtained by fitting the correlation
function, as in the case of the source radius. On the other hand,
there may be production mechanisms other than hadronization that
introduce some degree of incoherence. The study of BEC in particular
decays may help to disentangle the production processes involved.\\

%%%%%%%%%%%%%%%%

\section{Effects on the Dalitz plot distribution}

%%%%%%%%%%%%%%%%

Figure 2 shows the Dalitz plot distribution and its projections on $X$ and $Y$,
as  defined in Eq. (1), for the
decay $K^+ \rightarrow \pi^+ \pi^0 \pi^0$. The histograms show the projections
with (full line) and without (crosses) BEC. The scatter plot is the ratio of the
distribution with and without BEC, Where we took $\beta=25$ GeV$^{-2}$, which 
corresponds to $r=1$ fm.
Increasing $\beta$ (Eq. (10)) would make the effect more marked.
As mentioned above, the physical meaning of $\beta=4,9,16,25$ GeV$^{-2}$ 
can be viewed in terms of the source radius of $r \approx  0.4,0.6,0.8,1.0$ fm,
respectively.
The true value, however, should be extracted from experimental data measuring 
BEC in these kind of decays.\\ 

In what follows, we will show the plots with $\beta=7$ GeV$^{-2}$, {\it i.e.}, 
$r \approx 0.5$ fm, which is motivated by the electromagnetic radius of the $K^+$. 

Ignoring electromagnetic interactions in the final state, the distributions
for the modes (2), (3) and (4) would be identical.
The correction for this effect could be carried out dividing the phase space by the
Gamow coefficients, as was, done in Ref. \cite{ford}. Here, we try to focus on the
effect of two-particle interference, and will not introduce this correction.
In Ref. \cite{prep}, we will present a complete anlysis, including the interplay of
BEC and final-state electromagnetic interactions.\\

Figure 3-5 show the amplitudes $\mid M \mid ^2$ {\it vs.} $X$, as obtained 
using PDG values  \cite{pdg} (Table 1) for the different decay processes. The band
represents the uncertainty in  the parameters.
The lines are the amplitudes one should observe, after correction for 
BEC, for $\beta=7$ GeV$^{-2}$ and $\lambda=0.25, 0.5, 0.75$.

The bands in Figures 3-5 show the errors according to Table 1 values.
These curves are corrected and the corresponding errors are propagated.
The corrected curves with errors are then fitted
with the function of Eq. 1.  In this way we mimic the procedure
during data analysis and observe the change in the values of the parameters
as they would be observed by the experimentalist.\\
As one can see, the interference of identical particles in the final state 
changes the values  of $g$ and $h$.

In the decay $K_L \rightarrow \pi^0 \pi^0 \pi^0$, all the pion pairs are weigthed
according to Eq. (12). To good extension, the effect cancels
out, and the Dalitz plot remains almost unchanged. Figure 6 shows the the amplitude
using the PDG values \cite{pdg} (Table 1) for this decay. The scale in this plot
has been expanded to make the effect visible. In this case the fit is done with
a quadratic form without linear terms.\\

%%%%%%%%%%%%%%%%

\section{Discussion and Conclusions}

%%%%%%%%%%%%%%%%

The measured amplitud $\mid M \mid ^2$ as a function of $X$ changes its shape 
once BEC among identical particles in the final state products is taken into
account. The different curves shown in Fig. 4 represent the modified amplitude as a
function  of $X$  for a  pion source-radius of $r=0.5$ fm,  and a coherence
factor of $\lambda =0.25, 0.5, 0.75$.
The true values for $\beta$ and $\lambda$ must be extracted from data.
It may be that $\lambda$ and/or $\beta$ are  smaller than this, thereby reducing the
impact on the amplitude.\\
        
The Particle Data Group~\cite{pdg} gives the average values for $g$, $h$ and $k$
(see Eq. (1)). These are shown in Table 1.\\

\begin{table}
\centering
\begin{tabular}{lccc}
\hline
        Decay mode      & $g$   & $h$   & $k$ \\
\hline
$K^+\to \pi^+\pi^+\pi^-$ & -0.2154$\pm$0.0035 & 0.012$\pm$0.008  & -0.0101$\pm$0.0034 \\
$K^-\to \pi^-\pi^-\pi^+$ & -0.217$\pm$0.007   & 0.010$\pm$0.006  & -0.0084$\pm$0.0019 \\
$K^\pm\to \pi^\pm\pi^0\pi^0$ & 0.652$\pm$0.031 & 0.057$\pm$0.018
&0.0197$\pm$0.0045$\pm$0.0029 \\
$K_L^0\to \pi^+\pi^-\pi^0$   & 0.678$\pm$0.008 & 0.076$\pm$0.006 & 0.0099$\pm$0.0015 \\
$K_L^0\to \pi^0\pi^0\pi^0$   & ---             & -0.0033$\pm$0.0011$\pm$0.0007   & --- \\
                             & --- & (-0.0061$\pm$0.0009$\pm$0.0005)$^{\dagger}$ & --- \\
\hline
\end{tabular}
\caption{Average values of Dalitz slope parameters for $K\to 3\pi$ decays taken 
from the PDG \cite{pdg}. The NA48 collaboration \cite{new} has recentely published
the quadratic slope parameter for the decay $K_L^0\to \pi^0\pi^0\pi^0$. The result is
included here indicated with $^{\dagger}$. }
\label{pdgvalues}
\end{table}

After correcting the  $g$ and $h$ parameters for BEC, 
by fitting $1+gX+hX^2$ to the corrected X distribution - shown in Figs. 3,4 and 5,
one obtains the values given in Table 2.

\begin{table}
\centering
\begin{tabular}{lcc}
\hline
        Decay mode      & $g$ (BEC corrected)   & $h$ (BEC corrected) \\
\hline
$K^+\to \pi^+\pi^+\pi^-$     & -0.1744$\pm$0.0027   &  0.001$\pm$0.003 \\
$K^-\to \pi^-\pi^-\pi^+$     & -0.176$\pm$0.003     & -0.0008$\pm$0.0031 \\
$K^\pm\to \pi^\pm\pi^0\pi^0$ &  0.692$\pm$0.005     &  0.080$\pm$0.006 \\
$K^\pm\to \pi^0\pi^0\pi^0$ &          ------        & -0.0052$\pm$0.0023 \\
\hline
\end{tabular}
\caption{Central values of the BEC-corrected Dalitz slope parameters for $K\to 3\pi$ decays.
The error is the result of the fit to the corrected curves once the errors are propagated
in the correction procedure.}
\label{ourvalues}
\end{table}

\noindent
According to the $\mid \Delta I \mid = 1/2$ rule, one expects
\begin{equation}
  \frac{g(K^\pm\to \pi^\pm\pi^0\pi^0)}{g(K^\pm\to \pi^\pm\pi^\pm\pi^\mp)}
  = \frac{g(K_L^0\to \pi^+\pi^-\pi^0)}{g(K^\pm\to \pi^\pm\pi^\pm\pi^\mp)}
  = -2
\end{equation}
and
\begin{equation}
  \frac{g(K_L^0\to \pi^+\pi^-\pi^0)}{g(K^+\to \pi^+\pi^0\pi^0)} = 1.
\end{equation}

\noindent
Using the PDG values of Table \ref{pdgvalues}, for $g$ one obtains
\begin{eqnarray*}
  \frac{g(K^\pm\to \pi^\pm\pi^0\pi^0)}{g(K^- \to \pi^-\pi^-\pi^+)}
  &=& -3.0046 \pm 0.1726 \\[.2cm]
  \frac{g(K_L^0\to \pi^+\pi^-\pi^0)}{g(K^- \to \pi^- \pi^- \pi^+)}
  &=& -3.1244 \pm 0.1073 \\[.2cm]
  \frac{g(K_L^0\to \pi^+\pi^-\pi^0)}{g(K^+\to \pi^+\pi^0\pi^0)}
  &=& 1.0399 \pm 0.0509
\end{eqnarray*}
Using the values of Table \ref{ourvalues}, these become:
\begin{eqnarray*}
  \frac{g(K^\pm\to \pi^\pm\pi^0\pi^0)}{g(K^- \to \pi^- \pi^- \pi^+)}
  &=& -3.968 \pm 0.069 \\[.2cm]
  \frac{g(K_L^0\to \pi^+\pi^-\pi^0)}{g(K^- \to \pi^- \pi^- \pi^+)}
  &=& -3.887 \pm 0.075 \\[.2cm]
  \frac{g(K_L^0\to \pi^+\pi^-\pi^0)}{g(K^+\to \pi^+\pi^0\pi^0)}
  &=& 0.980 \pm 0.014
\end{eqnarray*}

The ratios change by taking into account a phase space distortion due to
the interference of identical particles in the final state.
We have given only a general trend based on parameter values motivated 
by the electromagnetic structure of the kaon. A more precise statement
would be possible once the experimentally extracted values of $\beta$ and $\lambda$
are used in the analysis.
One can see that the ratios given by Eq. (13) are very sensitive to a distortion
of phase space, while that of Eq. (14) is not. Nevertheless  although the size 
of the effect  may  be small, is clear that the observed violation of  
the $\mid \Delta I \mid = 1/2$ rule can be affected by 
BEC, and this should be taken into account in the anlaysis of data.

\section{Acknowledgement}

We thank CONACyT (M\'exico) for financial support.

\newpage

\section*{Figure Captions}
 
\begin{itemize}
\item 
[Fig. 1:] Diagram representing the Kaon decay process. The bubble represents
the space time region where pions are produced.

\item
[Fig. 2:] Dalitz plot distribution and its projections for the decay 
$K^+ \rightarrow \pi^+ \pi^0 \pi^0$. 
The histograms represent the distribution with (full line) and without (crosses)
a contribution from BEC. The distributions have been normalized to 
$\mid M^2 \mid = 1$ at $X=0$ and $Y=0$.
The scatter plot shows the ratio of the distribution with and without BEC with
parameters of the BEC simulation being $\beta=25$ GeV$^{-2}$ and $\lambda=0.5$.

\item 
[Fig. 3:] Amplitude $\mid M \mid ^2$ {\it vs.} $X$ for the decay 
$K^+ \rightarrow \pi^+ \pi^+ \pi^-$, as obtained from
experiment \cite{pdg} (shaded band) and the central value of the amplitude 
that one would obtain  (solid, shaded and dotted, lines) after correction for 
BEC effects with  $\lambda=0.25, 0.5, 0.75$. The band represents the experimental
uncertainties on the average values.

\item
[Fig. 4:] Amplitude $\mid M \mid ^2$ {\it vs.} $X$ for the decay
$K^- \rightarrow \pi^- \pi^- \pi^+$, as obtained from 
experiment \cite{pdg} (shaded band) 
and the central value of the amplitude that one would obtain 
(solid, shaded and dotted, lines)
after correction for BEC effects with $\lambda=0.25, 0.5, 0.75$.
The band represents the experimental uncertainties
on the average values.

\item
[Fig. 5:] Amplitude $\mid M \mid ^2$ {\it vs.} $X$ for the decay
$K^+ \rightarrow \pi^0 \pi^0 \pi^+$, as obtained from
experiment \cite{pdg} (shaded band) and the central value of the amplitude one would
obtain (solid, shaded and dotted, lines) after correction for BEC 
effects with $\lambda=0.25, 0.5, 0.75$.
The band represents the experimental uncertainties
on the average values.

\item
[Fig. 6:] Amplitude $\mid M \mid ^2$ {\it vs.} $X$ for the decay
$K^+ \rightarrow \pi^0 \pi^0 \pi^0$, as obtained from
experiment \cite{pdg} (shaded band) and the central value of the amplitude one would
obtain (solid, shaded and dotted, lines) after correction for BEC effects with $\lambda=0.25,
0.5, 0.75$. The band represents the experimental uncertainties on the average values.

\end{itemize}

\newpage

\begin{figure}[ht!]
\begin{center}
\includegraphics[width=\textwidth]{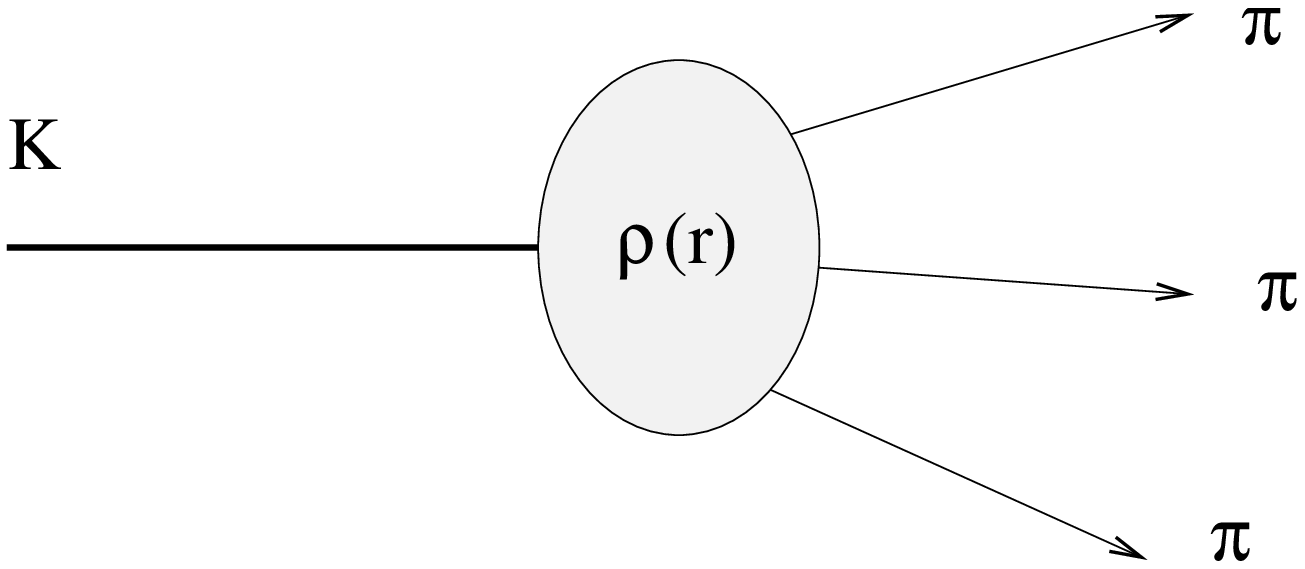}
\caption{ }
\end{center}
\end{figure}

\newpage

\begin{figure}[ht!]
\begin{center}
\includegraphics[height=.9\textheight]{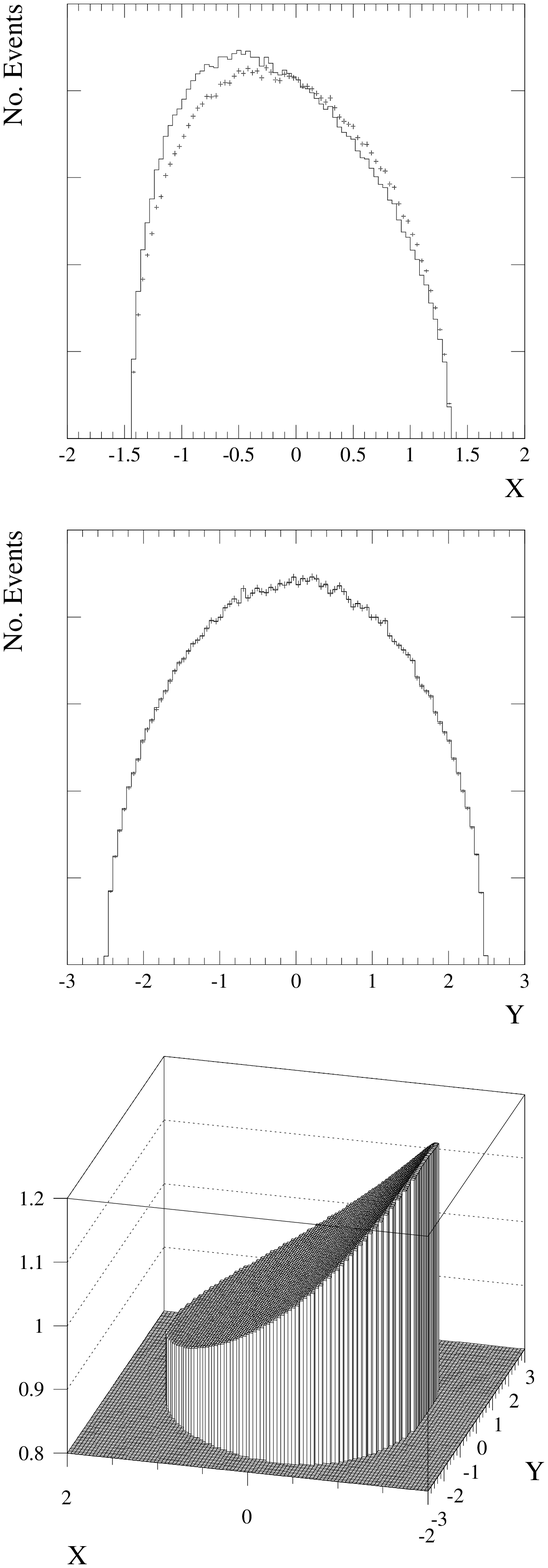}
\caption{ }
\end{center}
\end{figure}

\newpage

\begin{figure}[b]
\includegraphics[width=\textwidth]{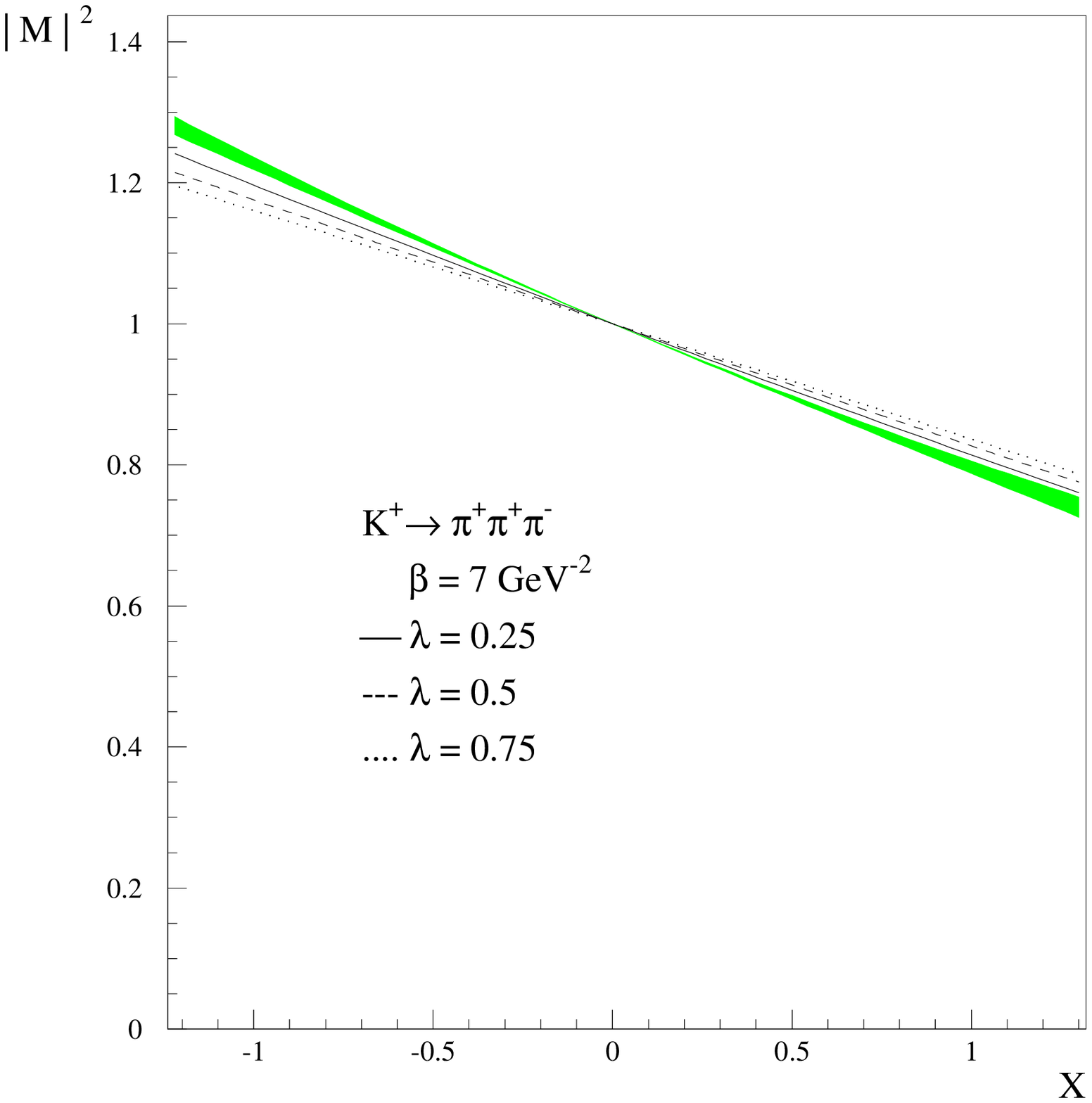}
\caption{ }
\end{figure}

\newpage

\begin{figure}[b]
\includegraphics[width=\textwidth]{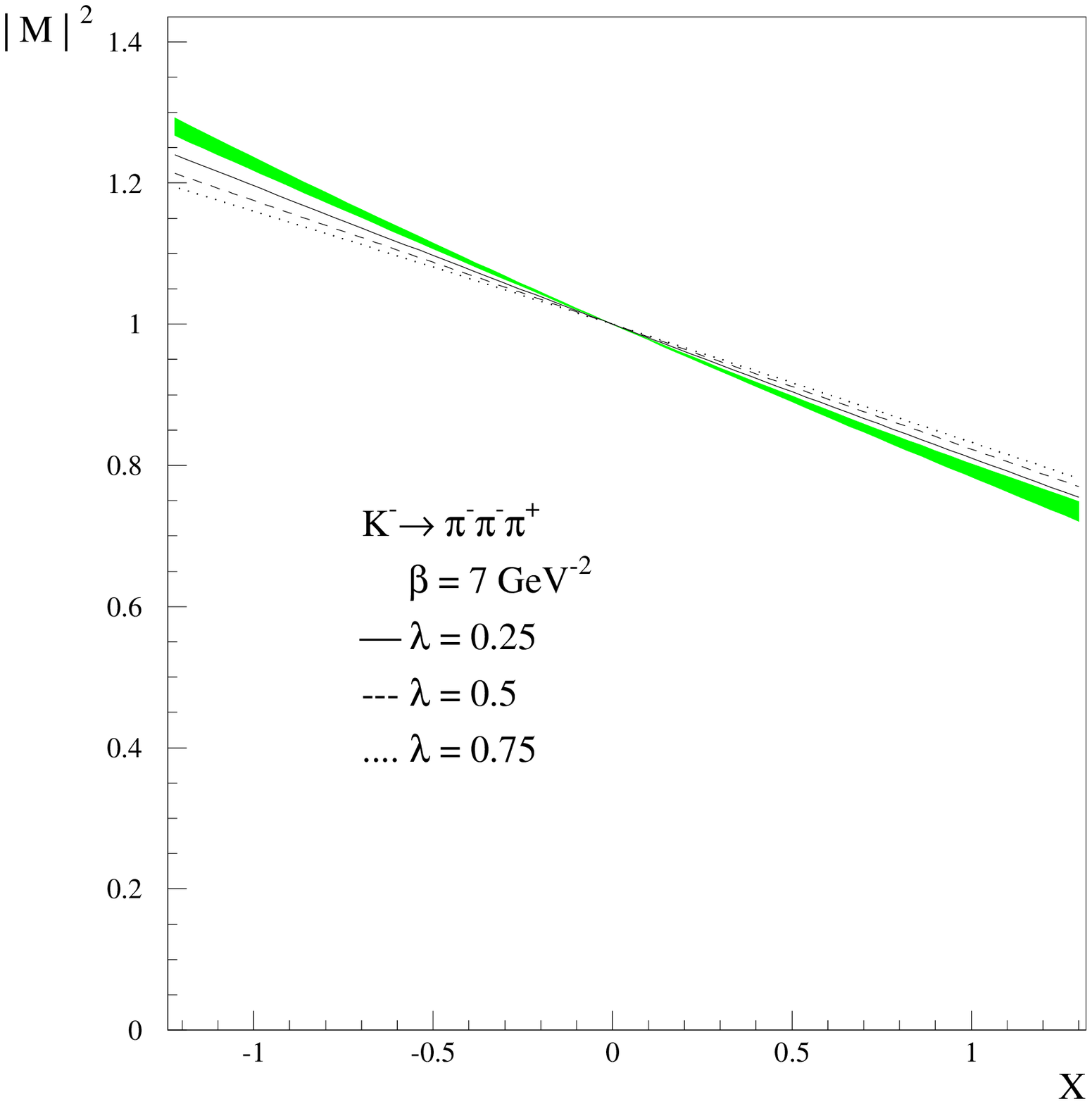}
\caption{ }
\end{figure}

\newpage

\begin{figure}[b]
\includegraphics[width=\textwidth]{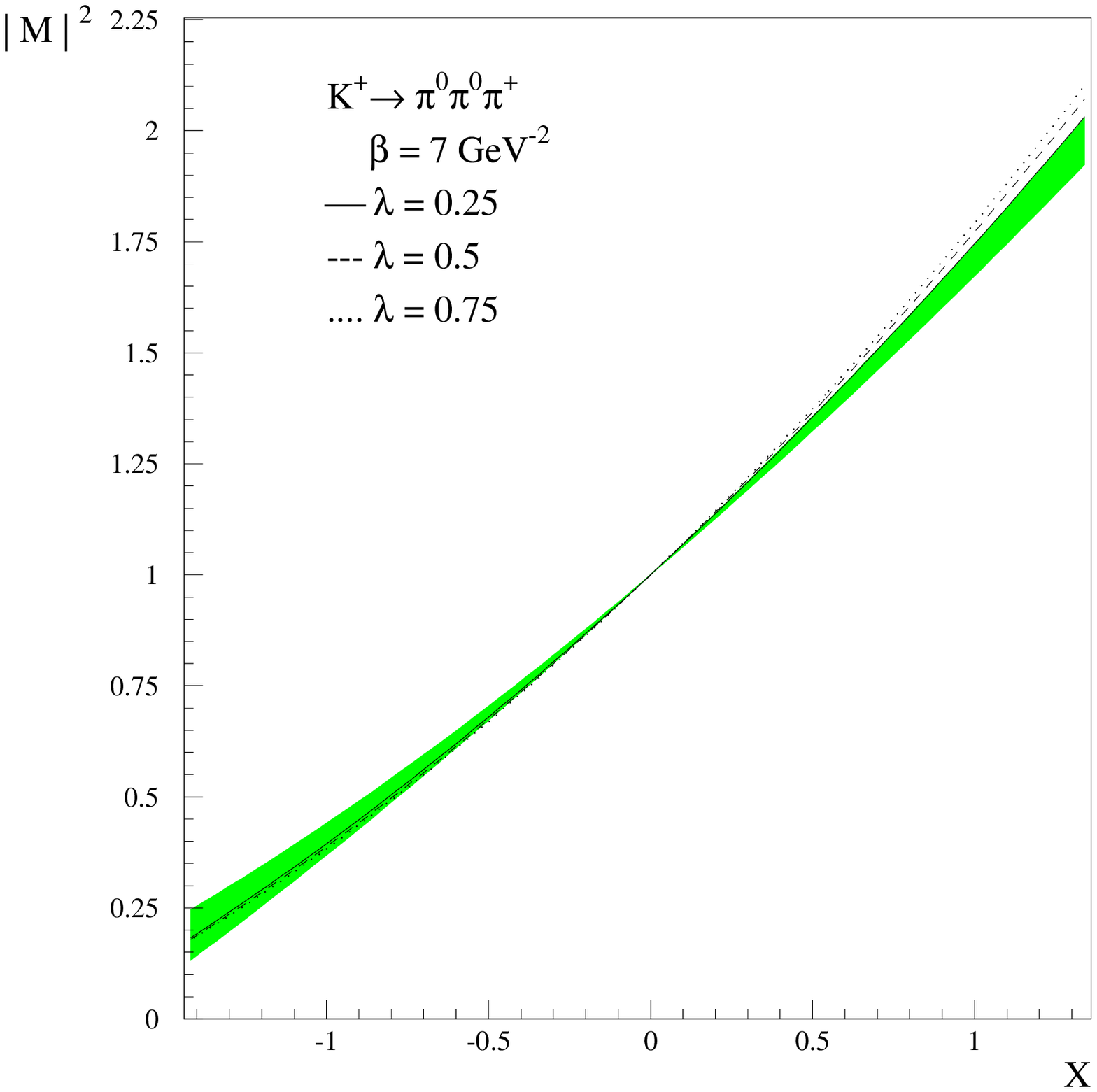}
\caption{ }
\end{figure}

\newpage

\begin{figure}[b]
\includegraphics[width=\textwidth]{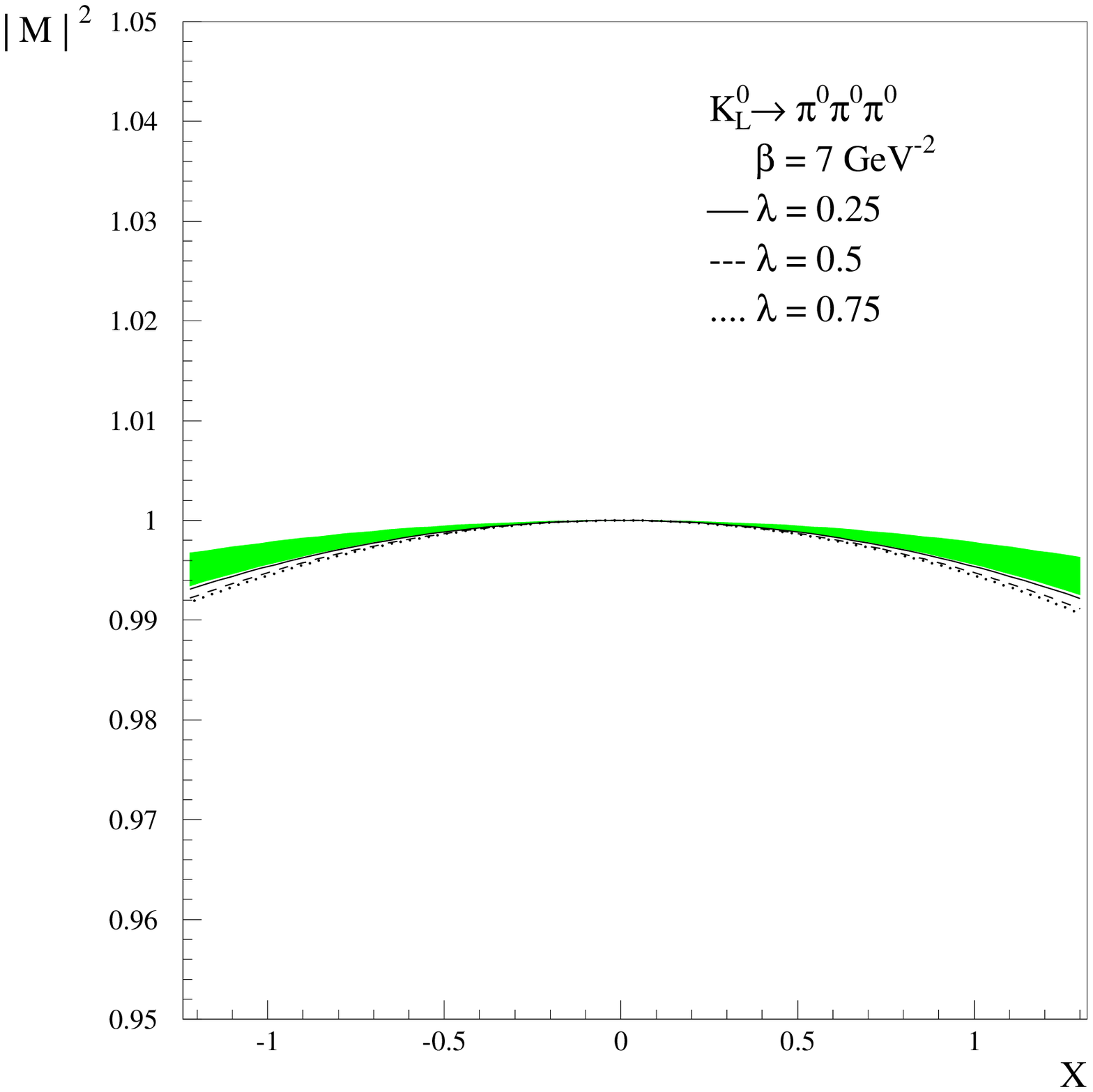}
\caption{ }
\end{figure}


\newpage


\begin{thebibliography}{99}

\bibitem{weinberg} S. Weinberg, {\it Phys. Rev. Lett.} {\bf 4} (1960)87 

\bibitem{pdg} Particle Data Group, Review of Particle Physics;\\
{\it Eur. Phys. J.} {\bf C15} (2000)1 

\bibitem{new} S. Somalwar {\it et al.}, {\it Phys. Rev. Lett.} {\bf 68}
(1992)2580\\
NA48 Collaboration (A. Lai {\it et al.}), {\it Phys.Lett.}, {\bf B515} (2001)261

\bibitem{batusov} V. Y. Batusov {\it et al.}, {\it Nucl. Phys.} 
{\bf B516} (1998)3

\bibitem{theory} J. Kambor, J. Missimer and D. Wyler, {\it Phys. Lett.}
{\bf B261} (1991)496\\
J. Kambor, {\it et al.} {\it Phys. Rev. Lett.}
{\bf 68} (1992)1818\\
A. Belkov, {\it et al.} {\it Int. J. Mod. Phys.} {\bf A7} (1992)4757\\
G. D'Ambrosio, {\it et al.} {\it Phys. Rev.} {\bf D50} (1994)5767

\bibitem{gglp} Goldhaber G., Goldhaber S., Lee W., Pais A., \\
{\it Phys. Rev.} {\bf 120} (1960)300

\bibitem{cuautle}  E. Cuautle, G. Herrera, {\it Phys. Lett.}
{\bf B434} (1998)153

\bibitem{residual} G.D. Lafferty, {\it Z. Phys.} {\bf C60} (1993)659

\bibitem{bialas} A. Bialas and A. Krzywicki, {\it Phys. Lett.}
{\bf B354} (1995)134

\bibitem{sjostrand} L. Lonnblad and T. Sjostrand, {\it Phys. Lett.}
{\bf B351} (1995)293

\bibitem{flow} Phuong Mai Dinh, Nicolas Borghini, Jean-Yves Ollitrault
{\it Phys. Lett.} {\bf B477} (2000)51

\bibitem{radius}  S.R. Amendolia {\it et al.}, {\it Phys. Lett.}{\bf B178}
(1986)435

\bibitem{us} R. Hern\'andez and G. Herrera, {\it Phys. Lett.} {\bf B332} 
(1994)448\\
A. Gago and G. Herrera, {\it Mod. Phys. Lett.} {\bf A10}(1995)1435

\bibitem{boal}D.H.Boal, C.K.Gelbke, B.K.Jennings, {\it Rev. Mod. Phys.} {\bf 
62} (1990)553

\bibitem{prep} M.I. Martinez and G. Herrera, in preparation

\bibitem{ford}W. T. Ford {\it et al.}, {\it Phys. Lett.} {\bf 38B}(1972)335

\end{thebibliography}
\end{document}